\documentclass[fleqn,usenatbib]{mnras}

\usepackage{newtxtext,newtxmath}
\usepackage[T1]{fontenc}
\usepackage{ae,aecompl}

\usepackage{hyperref}

\usepackage{graphicx}	% Including figure files
\usepackage{amsmath}	% Advanced maths commands
\usepackage{amssymb}	% Extra maths symbols

\title[FRB Death Star]{Fast Radio Bursts from Terraformation}

\author[A. Yalinewich et al.]{
A. Yalinewich,$^{1}$\thanks{E-mail: almog.yalin@gmail.com}
M. Rahman,$^{2}$,
A. Obertas$^{3,1}$ and
P. C. Breysse$^{1}$
\\
$^{1}$Canadian Institute for Theoretical Astrophysics, 60 St. George St., Toronto, ON M5S 3H8, Canada\\
$^{2}$Dunlap Institute for Astronomy and Astrophysics, University of
Toronto, 50 St.  George Street, Toronto, ON, M55 3H4, Canada\\
$^{3}$Department of Astronomy and Astrophysics, University of
Toronto, 50 St.  George Street, Toronto, ON, M55 3H4, Canada
}

\date{Accepted XXX. Received YYY; in original form ZZZ}

\pubyear{2018}

\begin{document}
\label{firstpage}
\pagerange{\pageref{firstpage}--\pageref{lastpage}}
\maketitle

\begin{abstract}
Fast radio bursts (FRBs) are, as the name implies, short and intense pulses of radiation at wavelengths of roughly one metre. FRBs have extremely high brightness temperatures, which points to a coherent source of radiation. The energy of a single burst ranges from $10^{36}$ to $10^{39}$ erg. At the high end of the energy range, FRBs have enough energy to unbind an earth-sized planet, and even at the low end, there is enough energy to vaporise and unbind the atmosphere and the oceans. We therefore propose that FRBs are signatures of an artificial terraformer, capable of eradicating life on another planet, or even destroy the planet entirely. The necessary energy can be harvested from Wolf-Rayet stars with a Dyson sphere ($\sim 10^{38}$ erg s$^{-1}$) , and the radiation can be readily produced by astrophysical masers. We refer to this mechanism as Volatile Amplification of a Destructive Emission of Radiation (VADER). We use the observational information to constrain the properties of the apparatus. We speculate that the non-repeating FRBs are low-energy pulses used to exterminate life on a single planet, but leaving it otherwise intact, and that the stronger repeating FRB is part of an effort to destroy multiple objects in the same solar system, perhaps as a preventative measure against panspermia. In this picture, the persistent synchrotron source associated with the first repeating FRB arises from the energy harvesting process. Finally we propose that Oumuamua might have resulted from a destruction of a planet in this manner.
\end{abstract}


%%%%%%%%%%%%%%%%%%%%%%%%%%%%%%%%%%%%%%%%%%%%%%%%%%

%%%%%%%%%%%%%%%%% BODY OF PAPER %%%%%%%%%%%%%%%%%%

\section{Introduction}

Fast radio bursts (FRBs) are intense, and short flashes of radiation with a wavelength of around one metre whose origins are among the final frontier of astrophysics. Their large dispersion measure (hundreds to about two thousand pc cm$^{-3}$) implies the source is of extragalactic origin. Their intrinsic duration (after de-dispersion) is of the order of microseconds, and their typical flux is a few Jansky. Their inferred brightness temperature greatly exceeds the Kellerman limit \citep{Tsang2006TheSources} and implies a coherent emission mechanism. For a comprehensive discussion of the observed properties of FRBs we refer the reader to the \href{http://www.frbcat.org}{FRB catalogue} \citep{Petroff2016FRBCAT:Catalogue} and references therein. For a list of proposed mechanisms for FRBs we refer the reader to the \href{https://frbtheorycat.org/index.php/Main_Page}{FRB theory wiki} \citep{Platts2018ABursts}.

The vast majority of scenarios involve ``natural'' sources. One example for a model involving an ``artificial'' source is the light sail \citep{Lingam2017FastSails}. In this work we consider a more nefarious artificial source for FRBs. We propose that FRBs are signatures of an alien weapon of mass destruction, capable of vaporising an earth size planet. The required radiant energy at the right frequency range can be produced using astrophysical maser, composed primarily of volatile compounds. We refer to this apparatus as the Volatile Amplification of a Destructive Emission of Radiation (or VADER). 

The plan of the paper is as follows. In section \ref{sec:constraints} we discuss the theoretical constraints on the VADER system from the observations. In section \ref{sec:discussion} we discuss the results and their implications.

\section{Theoretical Constraints} \label{sec:constraints}

\subsection{Energy Budget} \label{sec:energy_budget}

So far, two FRBs have been localised. The first one is the first repeating FRB 121102  \citep{Spitler2016ABurst}. The isotropic equivalent energy for each burst is about $10^{39}$ erg. The other is FRB 171020 \citep{Mahony2018AFRB171020}. In this case the isotropic equivalent energy is considerably lower - about $10^{36}$ erg.

The binding energy of a terrestrial planet of mass $M_p$ and radius $R_p$ is roughly given by
%https://www.wolframalpha.com/input/?i=(gravitation+constant)*(earth+mass)%5E2%2F(earth+radius)
\begin{equation}
    U_b \approx 10^{39} \left(\frac{M_p}{M_{\oplus}}\right)^2 \left(\frac{R_p}{R_{\oplus}}\right)^{-1} \, \rm erg \, .
\end{equation}
Therefore, a repeating burst has enough energy to entirely unbind a terrestrial planet. The energy in the non repeating burst would suffice to unbind the atmosphere and the oceans on the surface of the planet. 

The minimum mass of the emitter can be estimated by assuming that each molecule emits a single photon
%https://www.wolframalpha.com/input/?i=(proton+mass)*(1e39+erg)%2F((1+GHz)*(planck+constant))%2F(solar+mass)
\begin{equation}
    M_e > 0.1 \frac{\mu}{m_p} \frac{E}{10^{39} \, \rm erg} \left(\frac{\nu}{1 \, \rm GHz}\right)^{-1} M_{\odot}
\end{equation}
where $E$ is the energy of the burst, $\mu$ is the mass of a single molecule, $m_p$ is the proton mass and $\nu$ is the frequency of the radiation. In principle, it is possible to increase the efficiency of the emitter by exciting multiple degrees of freedom. This increase in energy is bounded by the number degrees of freedom, and therefore cannot reduce the minimum mass by more than about an order of magnitude.

We note that this method of destroying a planet usually requires less energy than diverting the planet in the habitable zone toward the host star. This is because usually the orbital Keplerian velocity is larger than the escape velocity from the planet. Moreover, the biggest challenge with this approach is to get rid of the planet's orbital angular momentum.

\subsection{Maser Emission}

Masers are a well known source of coherent radiation in astrophysics \citep{Gray1999AstrophysicalMasers}. Astrophysical masers are primarily produced by molecules comprising volatile elements (with the exception of Silicon). The lowest frequency ever recorded for an astrophysical maser is about 700 MHz for a CH maser \citep{Ziurys1985DetectionCH}, and the highest frequency is about 3.4 THz, from a CO maser \citep{Storey1981Far-infraredOrion}. This is consistent with the non detection of  FRBs below 200 MH \citep{Sokolowski2018NoBursts}, while most detections are at or above 800 MHz. However, account for lower apparent frequencies, redshift can.

Each of the maser lines is extremely narrow, but if multiple lines are emitted simultaneously, then when observed with coarse enough frequency resolution, the spectrum may seem continuous.

We note that since many of the compounds found in molecular clouds are also present in planets' atmosphere and mantles, then the emitted energy will be readily absorbed rather than reflected from the target.

\subsection{Duration}
One of the properties of coherent emission is that it can produce short, intense and polarised pulses. As the density of excited molecules increase, the intensity of the radiation increases and the duration decreases. This effect is often referred to a Dicke's superradiance \citep{Dicke1954CoherenceProcesses}. It has been shown that Dicke's superradiance in astrophysical masers can account for the observed duration and energy of FRBs \citep{Houde2017ExplainingSuperradiance}.

Even if the pulse is shorter than what is observed, the signal will be broadened due to reflection of the radiation from a curved target. The typical light crossing time for an earth sized planet is:

\begin{equation}
    t_{lc} \approx 20 \frac{R}{R_{\oplus}} \, \rm ms
\end{equation}

If the radius of the maser beam is smaller than the size of the planet, then the duration of the observed pulse will be shorter. If the size of the beam is a factor of 5 smaller than the radius of the planet, then the spread in arrival times of photons to the surface of the planet will be similar to the observed duration of FRBs (a few microseconds). 

\subsection{Pump}
We propose that the persistent synchrotron source associated with the first repeating FRB \citep{Chatterjee2017AHost} is related to the energy source used for population inversion in the maser. The radio luminosity of the persistent source is of the order of $10^{38}$ erg/s, which is comparable with the bolometric luminosity of some Wolf Rayet stars \citep{Hainich2014TheCloud}. We therefore propose that this energy is harvested by a Dyson sphere \citep{Semiz2015DysonDwarfs, Osmanov2016OnPulsars}.

The most straightforward way to transport the energy from the Wolf Rayet star is to accelerate its stellar wind to relativistic particles in a magnetically collimated beam. It has been previously estimated that the magnetic field is of the order of $B \approx 10 \, \rm mG$ and the Lorentz factor of the electrons is of the order of $\gamma \approx 100$ \citep{Waxman2017OnFRBs}. The bolometric luminosity of each electron is given by $L_1 \approx c \sigma_t B^2 \gamma^2$ (where $\sigma_t$ is the Thompson cross section) and the number of electrons is roughly given by $\dot{M} d/ c m_p$ where $\dot{M}$ is the mass loss rate, $d$ is the distance between the WR star and the molecular cloud and $m_p$ is the mass of the proton. The total synchrotron luminosity of the beam is therefore
%https://www.wolframalpha.com/input/?i=(thompson+cross+section)*(speed+of+light)*(1e-6*erg%2Fcc)*(100)%5E2*(1e-3*solar+mass)%2F(proton+mass)%2F(erg%2Fsec)
\begin{equation}
    L_b \approx 6 \cdot 10^{38} \frac{\dot{M}}{10^{-3} M_{\odot}/\rm y} \left(\frac{\gamma}{100}\right)^2 \frac{d}{\rm pc} \, \rm erg/s \, .
\end{equation}
Hence we get that the synchrotron emission from the beam is comparable with the observed value. This means that the majority of energy harvested from the WR star is spent in transporting this energy to the molecular cloud. This could explain why the time between consecutive bursts is considerably longer the ideal charging time (i.e. the ratio between the energy of an individual burst and the luminosity of the synchrtron source, about one minute).  In other words, the energy needed to destroy a planet may be insignificant next to the power of the source.

One property that sets the first repeater from other FRBs is the exceptionally high rotation measure, roughly $10^{5}$ rad/meter$^2$, which requires a high magnetic field. This magnetic field could be the same magnetic field that confines the beam.

A persistent synchrotron source of a similar power has not been detected around another localised, but non repeating, FRB \citep{Mahony2018AFRB171020}. Moreover, rotation measure of other FRBs \citep[e.g.][]{Petroff2017ALatitude} are significantly lower than that of the repeater. For this reason we assume that the pump is turned off prior to triggering maser emission. The pump remains active in the case of the repeater because the maser has to fire multiple times. One reason to do so is in order to destroy not just a single planet, but multiple objects in the same solar system.  This may be especially necessary, as the technology level necessary to create the system we describe here would also allow the creation of planet- or moon-sized objects which are in fact artificial space stations, increasing the number of targets in a single system.

\section{Interstellar Debris}

Recently, a first interstellar object, dubbed Oumuamua, has been detected passing through our solar system \citep{Jewitt2017InterstellarTelescopes}. Oumuamua was detected by the telescope PAN-STARRS, which have been observing for about ten years before the detection. If this detection is typical, and an interstellar object the size of Oumuamua enters our solar system once per decade, then this would require every solar system to eject debris with a total mass of about $4 M_{\oplus}$, which is problematic with current models of planet formation \citep{Do2018InterstellarObjects}.

\begin{figure*}
\begin{minipage}[c]{0.67\textwidth}
\includegraphics[width=0.7\textwidth]{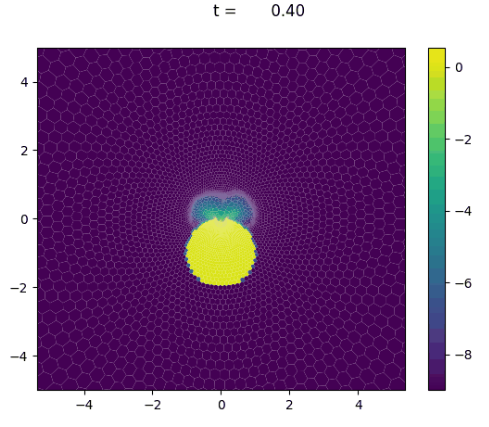}
\includegraphics[width=0.7\textwidth]{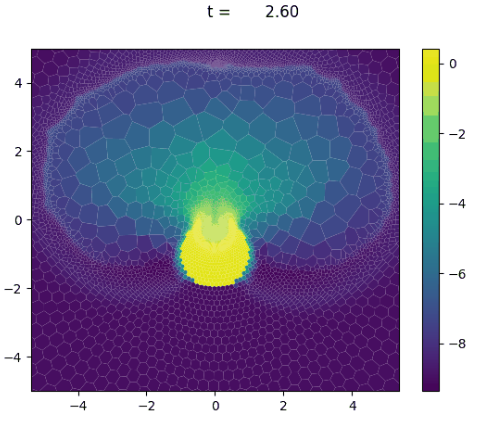}
\includegraphics[width=0.7\textwidth]{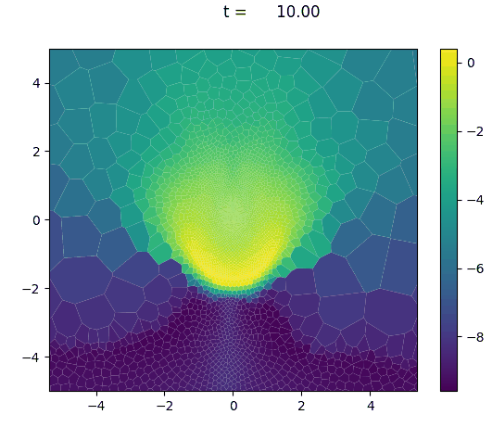}
\end{minipage}
\begin{minipage}[c]{0.3\textwidth}
\caption{
Log density snapshots from an numerical simulation of the passage of a shock wave in a terrestrial planet as a result of a deposition of radiative energy in the top part. The shock wave creates a substantial disturbance in the force of gravity holding the planet together and produces a wide debris field.
\label{fig:sim}
}
\end{minipage}
\end{figure*}

In this section we explore the prospect that Oumuamua is a part of the debris from a planet destroyed by the mechanism described in the previous section. A planet destroyed in this manner is expected to produce a wide debris field, as is illustrated by the numerical simulations presented in figure \ref{fig:sim}. If the escape velocity from the planet is greater than the Keplerian orbtial velocity, then the debris are guaranteed to leave the solar system. Stronger explosions can also expel debris out of the solar system from closer or less massive planets.

One of the peculiar features of Oumuamua, an object some already consider to be unnatural \citep{Bialy2018CouldAcceleration}, is the large variations in its light curve, which indicates a large aspect ratio \citep{Fraser2017The1I/Oumuamua}. Such a large aspect ratio cannot be readily produced in natural environments. Explosions, however, are known to produce irregularly shaped debris \citep{Baker1981AStructures}.

In order for a fragment to have reached us, the explosion had to have been relatively close, and therefore recent (in comparison to the age of the Galaxy). Travelling at roughly 20 km/s, an object would take about a million years to get to us if it were travelling at a straight line. Oumuamua came roughly from the direction of the constellation Lyra, which sports a number of relatively close exoplanets, like Kepler 37b \citep{Barclay2013AExoplanet}, Kepler 444 \citep{Campante2015ANPLANETS} and possibly Vega \citep{Harper1984OnVEGA}.  

If VADER mechanisms have been active in the Milky Way in the recent past, there exists a troubling possibility that such a system could be aimed at Earth. We do not calculate the probability of this, as the authors prefer not to be told the odds. The reader may get a bad feeling about this, as with current levels of technology, resistance to such a weapon would of course be futile.

\section{Discussion} \label{sec:discussion}

In this paper we discuss the possibility that fast radio bursts are signatures of an artificial device capable of destroying terrestrial planets. We propose that the device is based on maser emission in multiple spectral lines. Since such a molecular cloud is primarily composed of volatile elements, we refer to this mechanism as a Volatile Amplification of a Destructive Emission of Radiation (VADER). We show that the frequency range of FRBs is compatible with cosmologically redshifted maser lines, the energy of FRBs is comparable to the binding energy of terrestrial planets and that the de-dispersed FRB duration is comparable to the delay time from the reflection of a planar wave from a planet surface.

In this model, non-repeating FRBs are incidents where just a single planet in a particular solar system is destroyed, while the repeating FRBs are cases where multiple objects in the same system are destroyed. We note dynamical instabilities restrict the number of planets in the habitable zone to 5 or below \citep{Obertas2017TheStar}, whereas the number of repeated pulses from FRB121102 is close to 100 \citep{Zhang2018FastApproach}. This means that not only planets, but also moons and asteroids were destroyed. One reason to do so is to prevent panspermia \citep{MELOSH1988ThePanspermia, Horneck1994Long-termSpace}. The persistent synchrotron source and high rotation measure detected for the first repeating FRB are signatures of the pump energy source. We postulate that this energy source is a Wolf-Rayet star, harvested by a Dyson sphere. This energy is transferred to the molecular cloud by a relativistic particle beam.

As FRBs have been observed across a wide area of sky at extragalactic distances, this model implies that VADER mechanisms are active in numerous, widely separated galaxies.  From this, we can imply one of three things: either (a) many civilisations across the universe have independently developed this technology, possibly as a result of some kind of interstellar wars, (b) a single civilisation has existed for the multi-megayear timescales necessary to make the long trek between distant stars, or (c) some kind of ``hyperdrive" technology exists allowing for many-parsec journeys to be undertaken faster than relativity would allow (Solo, H., private communication).

To achieve higher efficiency, in terms of conversion of radiative energy to heat, the maser beam should be directed to a part of the planet surface that is rich in volatile elements and metals. On earth, such locations would be where there is fertile soil. For this reason, it could very well be that crop circles are target marks for such a weapon. Therefore, it could be that earth has been marked for destruction multiple times, perhaps to facilitate an intergalactic highway. We note, however, that this may not be the solution that you are looking for.

\section*{Acknowledgements}

We would like to thank George Lucas, Douglas Adams, Isaac Asimov, Gene Roddenberry, and Ridley Scott for their inspirational works.

%%%%%%%%%%%%%%%%%%%%%%%%%%%%%%%%%%%%%%%%%%%%%%%%%%



\bibliographystyle{mnras}
\bibliography{references} % if your bibtex file is called example.bib

%%%%%%%%%%%%%%%%%%%%%%%%%%%%%%%%%%%%%%%%%%%%%%%%%%

%%%%%%%%%%%%%%%%% APPENDICES %%%%%%%%%%%%%%%%%%%%%

%\appendix



%%%%%%%%%%%%%%%%%%%%%%%%%%%%%%%%%%%%%%%%%%%%%%%%%%

% Don't change these lines
\bsp	% typesetting comment
\label{lastpage}
\end{document}